\documentclass[apj]{emulateapj}
\usepackage{apjfonts}
\usepackage{epstopdf}

\def\wig#1{\mathrel{\hbox{\hbox to 0pt{%
          \lower.5ex\hbox{$\sim$}\hss}\raise.4ex\hbox{$#1$}}}}

\shorttitle{Clouds in GJ1214b}

\newcommand{\me}{$M_{\oplus}$}
\newcommand{\re}{$R_{\oplus}$}
\newcommand{\rsun}{$R_{\odot}$}
\newcommand{\rp}{$R_{\mathrm{p}}$}

\newcommand{\cp}{\citep}
\newcommand{\ct}{\citet}

\newcommand{\icarus}{Icarus} 
\newcommand{\fsed}{$f_{\rm sed}$} 
\newcommand{\nas}{Na$_2$S}
\newcommand{\rchi}{$\chi_{\rm red}^2$}
\newcommand{\kzz}{$K_{zz}$}

\slugcomment{Draft for ApJ}

\begin{document}

\title{Quantitatively Assessing the Role of Clouds in the Transmission Spectrum of GJ 1214b}

\author{Caroline V. Morley\altaffilmark{1,2}, Jonathan J. Fortney\altaffilmark{1}, Eliza M.-R. Kempton\altaffilmark{3}, Mark S. Marley\altaffilmark{4}, Channon Visscher\altaffilmark{5}, Kevin Zahnle\altaffilmark{4} }

\altaffiltext{1}{Department of Astronomy and Astrophysics, University of California, Santa Cruz, CA 95064; cmorley@ucolick.org}
\altaffiltext{2}{Harriet P. Jenkins Graduate Student Fellow} 
\altaffiltext{3}{Grinnell College} 
\altaffiltext{4}{NASA Ames Research Center} 
\altaffiltext{5}{Southwest Research Institute, Planetary Science Directorate} 

\begin{abstract}
Recent observations of the super-Earth GJ 1214b show that it has a relatively featureless transmission spectrum. One suggestion is that these observations indicate that the planet's atmosphere is vertically compact, perhaps due to a water-rich composition that yields a large mean molecular weight. Another suggestion is that the atmosphere is hydrogen/helium-rich with clouds that obscure predicted absorption features. Previous models that incorporate clouds have included their effect without a strong physical motivation for their existence. Here, we present model atmospheres of GJ 1214b that include physically-motivated clouds of two types. We model the clouds that form as a result of condensation in chemical equilibrium, as they likely do on brown dwarfs, which include KCl and ZnS for this planet. We also include clouds that form as a result of photochemistry, forming a hydrocarbon haze layer. We use a photochemical kinetics model to understand the vertical distribution and available mass of haze-forming molecules.  We model both solar and enhanced-metallicity cloudy models and determine the cloud properties necessary to match observations. In enhanced-metallicity atmospheres, we find that the equilibrium clouds can match the observations of GJ 1214b if they are lofted high into the atmosphere and have a low sedimentation efficiency (\fsed$=0.1$). We find that models with a variety of hydrocarbon haze properties can match the observations. Particle sizes from 0.01 to 0.25 \micron\ can match the transmission spectrum with haze-forming efficiencies as low as 1--5\%. 

\end{abstract}

 
\section{Introduction}

The transiting super-Earth GJ 1214b is the first planet discovered by the MEarth survey \cp{Charbonneau09} and is currently the only super-Earth that has been observed using transmission spectroscopy. The planet's mass and radius are 6.16$\pm$0.91\me\ and 2.71$\pm$0.24\re\ respectively \cp{Anglada-Escude13}, giving it a low bulk density of 1.68 g cm$^{-3}$. This density is consistent with either a water-rich planet or planet with a dense iron/rock core and hydrogen/helium envelope \cp{Nettelmann11, Seager07, Rogers10}. \ct{Rogers10} proposed three general mechanisms by which GJ 1214b may have accumulated its atmosphere. The planet may have accreted a hydrogen/helium envelope from the stellar nebula, outgassed a hydrogen envelope from a rocky planet, or contain a high water content in the interior with a hydrogen-depleted, water-rich envelope. \ct{Nettelmann11} argue that the water-rich hypothesis would require unreasonably large bulk water-to-rock ratios, suggesting that the atmosphere must be at least partially composed of hydrogen and helium. By measuring the composition of GJ 1214b's atmosphere using transmission spectroscopy, we can potentially distinguish between these hypotheses. 

\subsection{Transmission spectroscopy}

During a transit, the light from the host star passes through the atmosphere of the transiting planet. Because the opacity of the atmosphere varies with wavelength, the radius of the planet will appear to vary with wavelength. The depth of features in the transmission spectrum scales as $N_H\times 2H R_p / R_* $, where $N_H$ (the number of scale heights probed) is set by the opacities involved ($\sim1-10$), $H$ is the atmospheric scale height, \rp\ is the planetary radius, and $R_*$ is the stellar radius \cp{Seager00, Hubbard01}. The scale height $H$ is inversely proportional to the mean molecular weight $\mu$ of the atmosphere. By measuring the depth of transit features, we probe the mean molecular weight of the atmosphere and can thus probe whether the atmosphere is H/He-rich ($\mu\sim2.3$) or a higher mean molecular weight H$_2$O ($\mu\sim18$) atmosphere \cp{Miller-Ricci09}. 

Cloud opacity, due to equilibrium and non-equilibrium processes, can be readily seen in the atmosphere of every planet and moon with an atmosphere in our solar system.  These include sulfuric acid clouds on Venus \cp{Prinn73}, water and carbon dioxide clouds on Mars \cp{Montmessin06, Whiteway09}, ammonia clouds on Jupiter \cp{Baines02}, ammonia and water clouds on Saturn \cp{Sromovsky83, Baines09, Sanchez-Lavega11}, methane clouds and tholin haze on Titan \cp{Brown10, deKok07}, and methane-derived clouds and hazes on Uranus \cp{Pollack87, Irwin07, Karkoschka09} and Neptune \cp{Hammel89, Max03, Gibbard03}. It has long been recognized that clouds could impact the transmission spectrum of transiting exoplanets as well \cp{Seager00,Brown01b,Hubbard01}. Furthermore, at the slant viewing geometry relevant for transmission spectroscopy, it has been suggested that long light path lengths through the atmosphere could lead even minor condensates to become optically thick, thereby obscuring gaseous absorption features \cp{Fortney05c}.

Transmission spectroscopy has been successfully used to probe the atmospheres of hot Jupiters, enabling the detection of atoms, molecules, and even clouds \cp[e.g.][]{Charbonneau02, Pont08, Sing08a}. For GJ 1214b, if the atmosphere is H/He-rich the features are predicted to change the planet's radius by $\sim$0.1\% which would be detectable with current instruments \cp{Miller-Ricci10}. If the atmosphere is instead water-rich with $\mu\sim18$, the features will be a factor of $\sim~10$ smaller and the spectrum could appear featureless at the observational precision of current instrumentation. 

\subsection{Observations of GJ 1214b's atmosphere}

Numerous observations of the transmission spectrum of GJ 1214b have been made from optical through near-infrared wavelengths from both ground and space. \ct{Bean10}, using the Very Large Telescope, found that the transmission spectrum is featureless between 0.78 and 1.0 \micron. \ct{Desert11c}'s broad-band photometric observations using the \emph{Spitzer Space Telescope} at 3.6 and 4.5 \micron\ showed a flat spectrum. The high resolution NIRSPEC spectrum from \ct{Crossfield11} also showed a featureless spectrum. \ct{Croll11}, contradicting the other measurements, found a deeper \emph{K}-band (2.2 \micron) transit using the Canada France Hawaii Telescope, consistent with the larger features of an H$_2$-rich atmosphere. \ct{Berta12} obtained a near-IR spectrum using Wide Field Camera 3 on the \emph{Hubble Space Telescope} and found a transmission spectrum consistent with a featureless specturm. \ct{deMooij12} observed the transit of GJ 1214b in the optical bands \emph{g}, \emph{r}, \emph{i}, \emph{I} and \emph{z} and near-infrared bands \emph{K$_{\rm s}$} and \emph{K$_{\rm c}$} and found all but the \emph{g}-band observation to be consistent with a featureless spectrum. The \emph{g}-band point has a slightly higher radius, possibly indicative of scattering. \ct{Murgas12} observed GJ 1214b using the Gran Telescopio Canarias with a narrow-band tunable filter at three bands: one centered on the line core of H$\alpha$ and two in the continuum, centered on either side. Their data are consistent with previous observations, but show a high intrinsic scatter. \ct{Fraine13} re-observe the transit of GJ 1214b with \emph{Spitzer} and in \emph{I+z} bands from the ground; their results are also consistent with a featureless spectrum or a water-vapor atmosphere, and find that their best-fitting model has a transit radius that increases into the optical, indicative of a scattering constituent in the upper atmosphere. 

We note that some of the observations disagree with each other to high significance, especially in the near-infrared \emph{K}-band. Impartially, we adopt the errors published in the literature and accept that no model will agree with all points. 

\subsection{Previous cloud and haze models of GJ 1214b}

\ct{Fortney05c} suggested that in the slant viewing geometry of transmission spectroscopy, minor condensates could have appreciable optical depth. These minor condensates and hazes would lead to weaker than expected or undetected gaseous absorption features. 

Several previous studies have included some of the effects of clouds in GJ 1214b's atmosphere. One method of including cloud opacity is to include an ad hoc opaque level at the pressure level in the atmosphere required to reproduce the observations, which represents an optically thick (at all wavelengths) cloud deck \cp[e.g.][]{Berta12}. \ct{Benneke12} developed a Bayesian retrieval method for super-Earths which can incorporate this type of opaque level to represent a cloud. In a more sophisticated cloud treatment, \ct{Howe12} incorporate a range of haze layers into GJ 1214b atmospheres. In each of their models, they include a haze composed of polyacetylene, tholin, or sulfuric acid. They test a range of different ad hoc number densities, particle sizes, and pressure levels for each cloud material. They find that a hydrogen-rich atmosphere with a haze layer is generally consistent with the observations, but cannot rule out a water-rich atmosphere. Their result serves as a useful parameter study, demonstrating that clouds within a hydrogen-rich atmosphere can match the observations. 

 \begin{figure}[ht]
 \center    \includegraphics[width=3.75in]{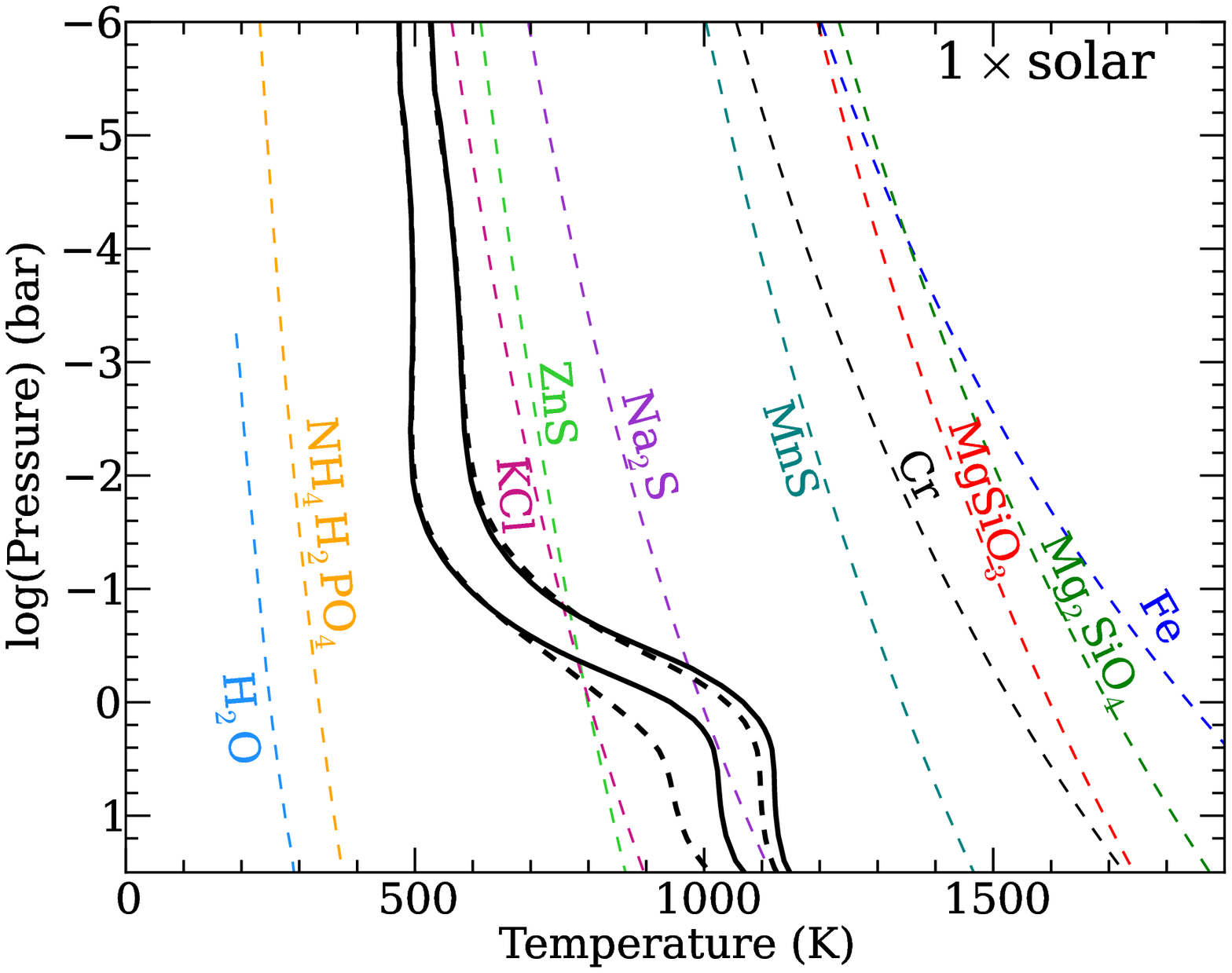}
\vspace{-0.5in}
 \center   \includegraphics[width=3.75in]{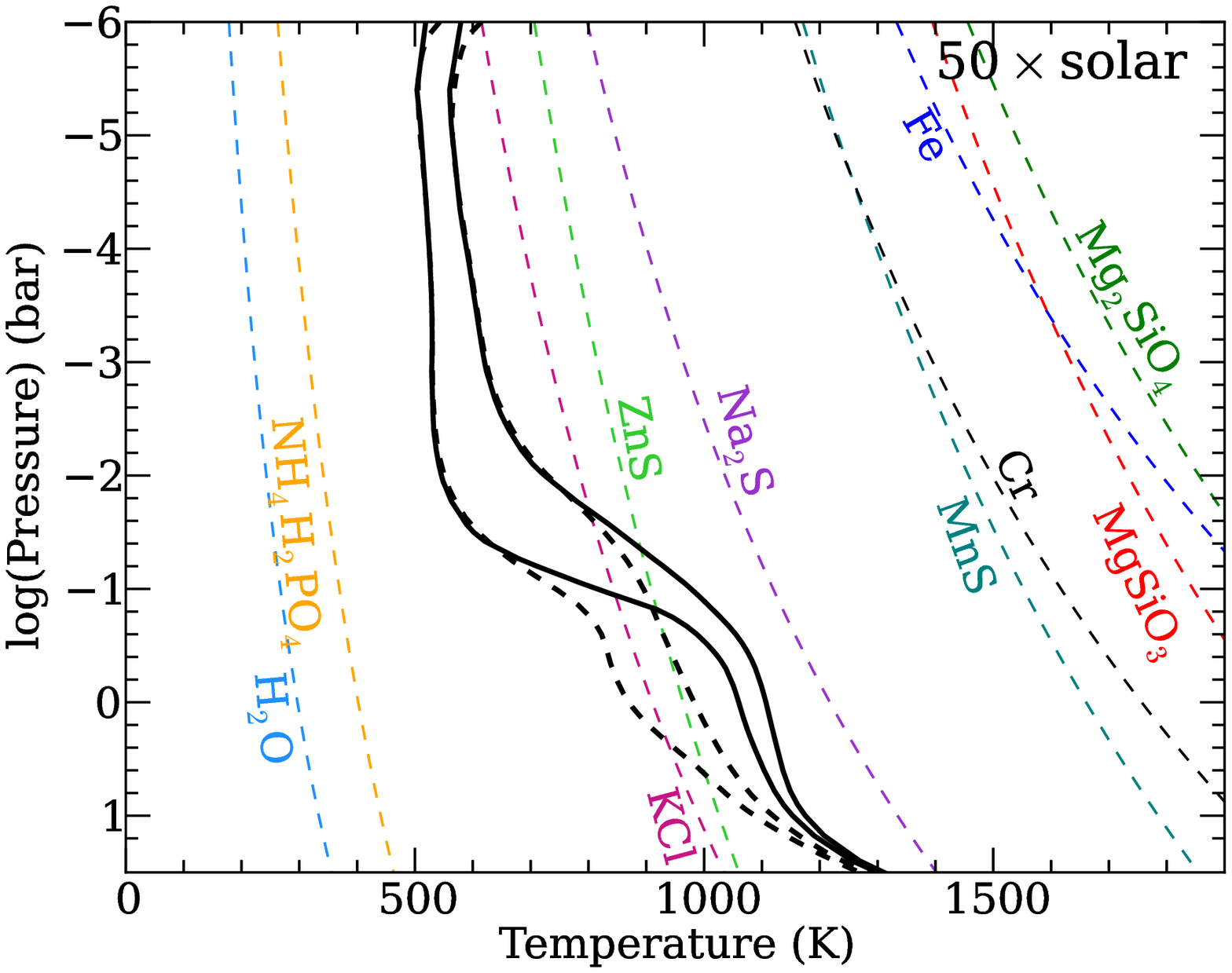}
  \caption{Pressure--temperature profiles of GJ 1214b with condensation curves. Top: solar composition models and condensation curves. Bottom: 50$\times$ solar models and condensation curves. Cloud-free \emph{P--T} profiles are shown as solid black lines; cloudy (KCl and ZnS clouds) models are shown as dashed lines. The cooler (left) models in each panel assume that the absorbed radiation from the star is redistributed around the entire planet, the warmer (right) ones assume that the radiation is redistributed over the dayside only. Condensation curves of all relatively abundant materials that will condense in brown dwarf and planetary atmospheres are shown as dashed colored lines.   See \S2.5 for a description of the models.}
\label{pt-profile}
\end{figure}

\subsection{Clouds from equilibrium and disequilibrium processes}

Previous cloud studies of GJ 1214b have invoked clouds as a way of matching the transmission spectrum observations, but these studies all lack a physical basis for choosing the cloud-forming material, the amount of cloud material, and the distribution of the cloud in the atmosphere; that is, a cloud layer could match observations, but no chemistry models were used to determine where clouds form and from what materials. While these studies do find that clouds can reproduce the observations, a remaining essential question is how plausible these clouds are given the conditions in the planet's atmosphere. In this study, we include two sets of physically-motivated clouds---based on two types of chemistry models---that are expected to form in the planet's atmosphere \cp{Kempton12}, and explore their effects in detail. 

The first set of clouds are those that form as a result of equilibrium chemistry. Equilibrium clouds have been extensively studied in brown dwarfs; silicate and iron clouds condense in L dwarfs \cp[e.g.][]{Tsuji96, Allard01,Marley02,Burrows06,Cushing08, Visscher10} and sulfide and chloride clouds condense in T dwarfs \cp{Lodders06, Visscher06, Morley12}. 

The other set of clouds we include form as a result of disequilibrium chemistry; we include a photochemically-produced haze layer. We follow the photochemical destruction of CH$_4$ and the corresponding creation of higher order hydrocarbons, with a photochemical model. Although we do not follow the photochemical pathways completely to haze formation, the model is used to determine the abundance and vertical distribution of haze precursors. 

Photochemical hazes form in the atmospheres of all of the solar system's giant planets \cp[e.g.][]{Gautier89}. While GJ 1214b is significantly warmer than any of these planets, it is cool enough that methane is still the most abundant carbon-bearing species \cp{Kempton12}. Due to its large UV photodissociation cross section, methane breaks apart in the upper atmosphere of irradiated planets and produces rich carbon chemistry in the atmosphere. Models that include UV dissociation of methane find that molecules such as C$_2$H$_2$, C$_2$H$_4$, C$_2$H$_6$, CH$_3$, HCN, and C$_6$H$_6$ exist in far greater abundance than would be expected from chemical equilibrium calculations \cp{Yung84, Zahnle09, Moses11, Kempton12}.

\subsection{Other approaches to cloud formation in brown dwarfs}
There are of course many ways to model clouds in planetary atmospheres \cp{Marley13}, and because we approach the problem by extending the models which have been used mainly for brown dwarf science, we will briefly describe the differences between the approaches here. For a much more extensive comparison, \ct{Helling08} review various cloud modeling techniques and compare model predictions for various cases. 

Many modeling groups apply the general assumptions of equilibrium chemistry that we apply here \cp[e.g.][]{Tsuji96, Allard01,Marley02,Burrows06,Cushing08, Visscher10}, though these models differ in the details of their approaches. A detailed comparison of true equilibrium condensation and cloud condensate removal from equilibrium can be found in \ct{Fegley94, Lodders06} and references therein.

Instead of assuming equilibrium chemistry to calculate cloud propertiers, Helling \& Woitke \cp{Helling06} follow tiny seed particles of TiO$_2$ in the upper atmospheres of brown dwarfs, which they assume to be advected from deeper layers, and follow the particles as sink downwards. As these seed particles fall through the atmosphere, they collect condensate material.  In \ct{Helling06} and
numerous follow on papers \cp{Helling08, Witte09, Witte11, deKok11} this group models the microphysics of grain growth given these conditions. They predict `dirty' grains composed of layers of varying condensates. This model has been applied to self-luminous giant planets and brown dwarfs but has not yet been applied to transiting super-Earths like GJ 1214b.

\section{Methods}

\subsection{Atmospheric composition}
\label{atmcomp}

The only planets in a similar mass range to GJ 1214b with well-characterized atmospheric compositions are Uranus and Neptune. Both planet's atmospheres are $\sim$ 50$\times$ solar abundance in carbon, mostly in the form of methane \cp{Fletcher10}. Other elements cannot easily be studied in those atmospheres because the planets are cold and most species are condensed into clouds below the visible atmosphere. 

Although it is well-established that planetary atmospheres in our own system have enhanced metallicities, the composition of exoplanet atmospheres is not yet well understood. In this analysis, we include solar composition models and 50$\times$ solar metallicity ([M/H]=1.7) models. The enhanced metallicity models are enhanced uniformly in all heavy elements. 
 \begin{figure*}[ht]
  \center    \includegraphics[width=6in]{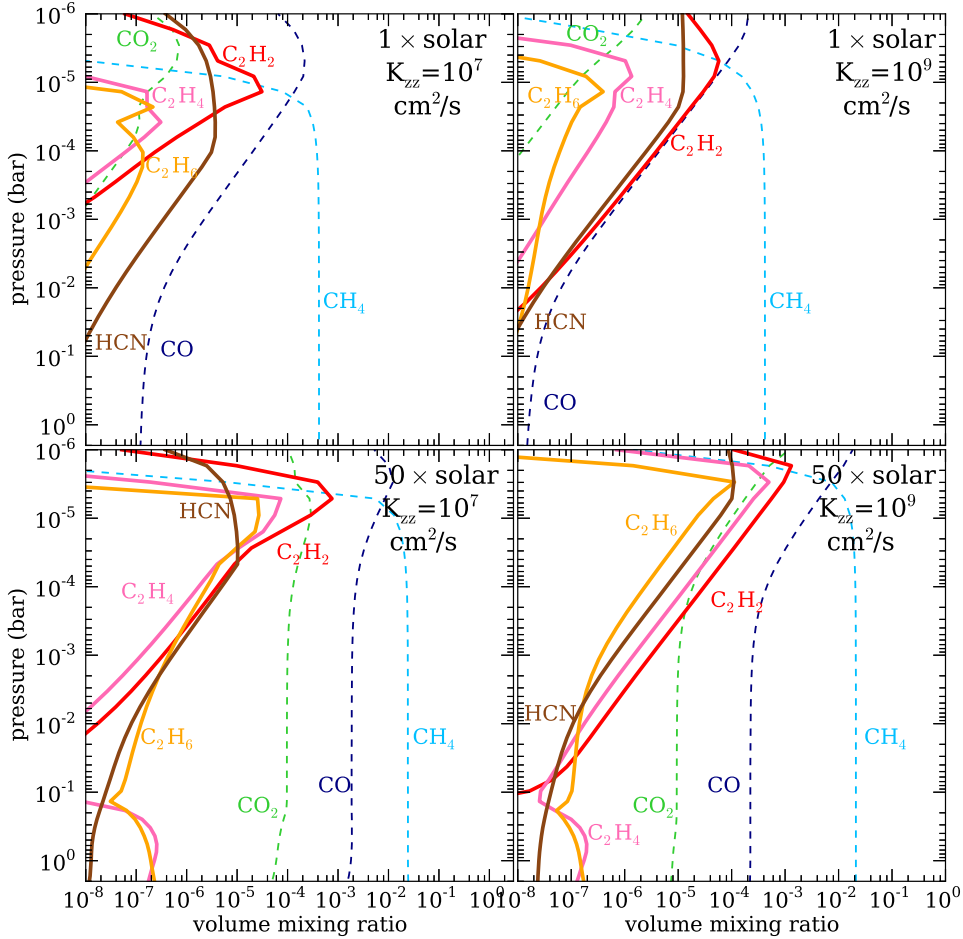}
 \caption{Results from photochemical calculations for C-bearing species at 1$\times$ (top) and 50$\times$ (bottom) solar metallicity. The volume mixing ratio at each pressure level of the atmosphere is shown for the major C-bearing species. The left and right panels shows the results using an eddy diffusion coefficient of \kzz$=10^7$ and \kzz$=10^9$ cm$^2$ s$^{-1}$, respectively. A fraction of the C$_2$H$_2$, C$_2$H$_4$, and C$_2$H$_6$ and HCN formed are assumed in this study to form the photochemical haze layer; CO, CO$_2$, and CH$_4$ do not readily form haze material. }
\label{photochem}
\end{figure*}

\subsection{Equilibrium cloud models}
\label{methods-eqclouds}

In GJ 1214b's atmosphere, assuming thermochemical equilibrium, a variety of substances will condense in the upper atmosphere. Figure \ref{pt-profile} shows pressure--temperature (\emph{P--T}) profiles of model atmospheres with hydrogen/helium rich compositions. The top panel shows solar composition models and condensation curves; the bottom panel shows 50$\times$ solar metallicity composition models. Cloud-free models are shown as solid lines and cloudy models are shown as dashed lines. The models used to calculate these profiles are discussed in Section \ref{atmmodel}. The condensation curves, shown as colored dashed lines, show where each element or molecule condenses assuming chemical equilibrium; the condensation curve represents the pressures and temperatures where the vapor pressure of a gas is equal to its saturation vapor pressure. To the left of the curve, we assume that vapor in excess of the saturation vapor pressure will form a condensate that settles toward the cloud base. In this approach, the cloud base forms where the \emph{P--T} profile meets the condensation curve, with some vertical extent above that point. The iron and silicate clouds form extremely deep in the atmosphere, as they do on Jupiter. \nas, ZnS, MnS, Cr, and KCl form higher in the atmosphere.

The opacity of \nas, ZnS, MnS, Cr, and KCl clouds have recently been included in T dwarf atmospheres by \ct{Morley12}, using the equilibrium condensation approach of \ct{Visscher06} within the framework of the well-established \ct{AM01} treatment for cloud formation and settling in L dwarf atmospheres. We use the methods developed and described in \ct{Morley12} to include the same clouds in a super-Earth atmosphere. 

The major differences between the \ct{Morley12} models for brown dwarf atmospheres and the models here are the irradiation of the planet by the host star and the enhanced metallicity of the atmosphere for some models. \ct{Morley12} published saturation vapor pressure and condensation curves for [M/H]$=-$0.5 and $+$0.5. For this study, a higher metallicity, [M/H]$=1.7$, was necessary. We calculated the condensation temperature for KCl and ZnS based on a model GJ 1214b \emph{P--T} profile. We found that, for this particular model atmosphere, the saturation vapor pressure and condensation curves were very close to the values we would have found by extrapolating the \ct{Morley12} vapor pressure and condensation curves. Since the differences due to the extrapolation are small for the gases in question, we adopted the same curves in this study.

To flatten transmission spectrum features in the near-IR, the clouds must be present and optically thick at \emph{slant} viewing geometry above $\sim$10$^{-3}$ bar. Only KCl and ZnS form that high in the atmosphere for this planet (see Figure \ref{pt-profile}), so for the models here, we include only the KCl and ZnS clouds. \nas\ will also form, but generally too deep to become optically thick at the low pressure levels probed in transmission spectra. We assume that the KCl and ZnS form into homogeneous, spherical particles, unlike the heterogeneous compositions that have been favored by \ct{Helling08b}. 

The particle sizes and vertical thickness of the cloud are calculated using the parametrized value \fsed\ in the \ct{AM01} framework. This value is equal to the ratio of the sedimentation velocity to the updraft velocity. A high sedimentation efficiency \fsed\ forms a cloud with large particles that settles into a thin layer; a low \fsed\ forms a more extended cloud with smaller particles. 

\subsection{Photochemistry models}

The photochemistry model used to generate the properties of the hydrocarbon haze layer is described in detail in \ct{Kempton12}. At the time of publication of \ct{Kempton12}, the UV spectrum of GJ 1214 had not been observed; instead, the study focused on two end-cases: a quiet M dwarf and an active M dwarf (AD Leo). Recently, \ct{France13} published a UV spectrum of the GJ 1214 host star. For the host star in the photochemistry model, we use this observed UV spectrum from 1150 to 3100 \AA\ and a PHOENIX model atmosphere spectrum with a stellar effective temperature of 3026 K and a stellar radius of 0.211 \rsun\ from 3100 to 10000 \AA\ \cp{Hauschildt99}.  We have calculated new solar metallicity and 50$\times$ solar metallicity models with the new spectrum. Figure \ref{photochem} shows the results of the photochemistry calculations for the two values of the eddy diffusion coefficient used in this study: \kzz= 10$^7$ and 10$^9$ cm$^2$ s$^{-1}$.

\subsection{Hydrocarbon haze}
\label{hydrocarbonhaze}

As described above, for the equilibrium clouds, our model parametrizes the cloud properties with a single value, \fsed; this model inherently assumes that the condensation of particles can be described by an equation for saturation vapor pressure which is an analytic function of atmospheric temperature and pressure. The formation of a hydrocarbon haze layer by polymerization is more complex and cannot be described simply and analytically in the same way as a function of temperature and pressure. The more complicated situation means that we cannot use the single parameter \fsed, but instead must calculate based on the photochemistry the amount of haze in each layer, and explore how changing parameters like particle size affects the transmission spectrum. 

\subsubsection{Forming soots from second-order hydrocarbons}

\begin{figure}[t]
    \includegraphics[width=3.75in]{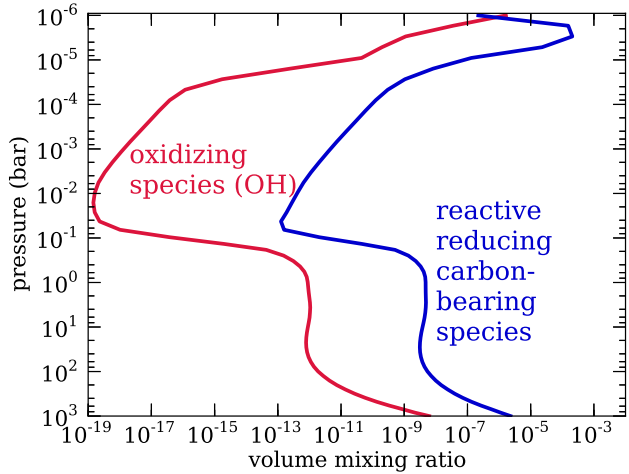}
 \caption{Comparison of reducing and oxidizing species for 50$\times$ solar, $K_{zz}$=10$^9$ cm$^2$ s$^{-1}$ photochemical model. The volume mixing ratio of the major oxidizing species (OH) and summed mixing ratio of all the major reactive reducing species (C$_2$H, C$_2$H$_3$, CH, CH$_2$, CH$_3$, CN) are plotted. There is significantly more reducing material at the pressure levels where we form hazes, so we assume that higher-order hydrocarbons will continue to grow to potentially form condensed hydrocarbon soot-like particles. } 
\label{testphotochem}
 \end{figure}

The highest order hydrocarbons produced by the \ct{Kempton12} and \ct{Zahnle09} models are the second-order hydrocarbons acetylene (C$_2$H$_2$), ethylene (C$_2$H$_4$), and ethane (C$_2$H$_6$). Higher-order hydrocarbon chemistry (e.g., >C$_3$H$_x$) in reducing, high-temperature, low-pressure planetary environments like GJ 1214b remains incompletely understood, and current photochemical and kinetics models (which generally derive reaction rates from combustion studies under much more oxidizing conditions) do not capture all possible chemical pathways for producing higher-order hydrocarbons that form soots in exoplanet atmospheres (e.g., see \ct{Moses11} discussion on C$_3$--C$_6$ chemistry). 

Because the hydrocarbon chemistry is truncated at C$_2$H$_x$, polymerization beyond C$_2$H$_x$ is not included. When conditions favor polymerization carbon will instead pool in C$_2$H$_x$ species, because longer carbon chains are not allowed. We can estimate how favorable conditions are for polymerization by comparing the quantities of reducing and oxidizing species in the atmosphere. If there are more oxidizing species (OH), oxidation by OH will inhibit hydrocarbon polymerization. If there are instead more reactive reducing species (including C$_2$H, C$_2$H$_3$, CH, CH$_2$, CH$_3$, CN), then hydrocarbon polymerization is not inhibited and is expected to continue at some rate (to date not well constrained by either experiments or kinetic theory). We calculate the amount of oxidizing and reducing material in each model atmosphere and determine that the amount of reducing material is larger---often many orders of magnitude larger---than the amount of oxidizing material at the pressure levels where soots are expected to form. Each soot precursor will therefore react many times with these reducing radicals before interacting with an OH molecule, growing progressively larger until it become involatile enough to condense to form a solid soot-like haze particle. Figure \ref{testphotochem} shows an example of this comparison for a photochemistry model with 50$\times$ solar composition and $K_{zz}$=10$^9$ cm$^2$ s$^{-1}$. This is not unexpected in a cool atmosphere like GJ 1214b's (T$_{\rm eff}\sim$550 K) with a relatively inactive host star (see \ct{Zahnle09} for more details).

Because of this propensity to polymerize, we assume that the second-order hydrocarbons C$_2$H$_2$, C$_2$H$_4$, and C$_2$H$_6$ and HCN in GJ 1214b's atmosphere continuously polymerize to form complex hydrocarbons like soot. We further assume that this process happens with the same constant efficiency in each layer of the atmosphere.  We treat this efficiency as a free parameter.  Hazes are thus most likely to form at altitudes where these soot precursors (C$_2$H$_2$, HCN) are produced in abundance via photochemical and thermochemical processes. The soot precursors are most favored when CH$_4$ is abundant, as is the case for GJ 1214b, and will be enhanced further for high C/O ratios \cp{Moses13}. 

\subsubsection{Calculating the hydrocarbon haze properties}

To determine the amount of material available to form hydrocarbon haze, we use the results from photochemistry models. These results give us the mixing ratio of each species at each pressure level in the atmosphere (see Figure \ref{photochem}). We calculate the number density of each species at each height in our model and multiply by the mean molecular weight of each species to calculate the mass density of C$_2$H$_2$, C$_2$H$_4$, C$_2$H$_6$, and HCN in each model layer. We sum the densities of these four species to find the total mass in soot precursors. A fraction of the total mass of soot precursors goes into forming the haze we model here: we multiply the total mass by our parametrized ``efficiency''---that is, the fraction of haze precursors that actually form haze particles---to find the mass of the haze particles in a given layer. For each layer, 

\begin{equation}
\label{fhaze-eq}
M_{\rm haze} = f_{\rm haze} \times (M_{\rm C_2H_2} + M_{\rm C_2H_4} + M_{\rm C_2H_6} + M_{\rm HCN})
\end{equation}

where $f_{\rm haze}$ is the prescribed efficiency, $M_x$ is the mass of material in each species within each model layer from the photochemical model, and $M_{\rm haze}$ is the calculated mass of haze particles in that layer. 

From the total mass of haze particles in each layer, we calculate how many particles form. We choose a mode particle size and establish a log-normal particle distribution; we calculate the number of particles by summing over the distribution for each of our chosen particle sizes.

We base our particle size distribution and physical properties on those found in experiments of soots on Earth. For example, \ct{Kim99} finds that diesel soot particles can have mode particle sizes between 0.05 and 0.5 \micron\ with a relatively log-normal distribution around the mode. We use an average material density from \ct{Slowik04} of 2.0 g cm$^{-3}$; note that while soots often form as low-density fluffy aggregates on Earth, we use the density only to calculate the number of particles formed, so the density of the solid soot material must be used. We use soot optical properties (the real and imaginary parts of the refractive index) tabulated in the software package OPAC (Optical Properties of Aerosols and Clouds) \cp{Hess98}, which we linearly extrapolate for wavelengths longer than 40 \micron. The extrapolation affects the spectrum negligibly between 40 and 230 \micron. 

When calculating the transmission spectrum, equilibrium chemistry abundances are used. \ct{Kempton12} showed that the disequilibrium abundances of carbon and nitrogen species will change the calculated spectrum very slightly, but will not change the overall shape of the spectrum.

\subsection{Atmosphere model}
\label{atmmodel}

The equilibrium cloud code is coupled to a 1D atmosphere model that calculates the pressure--temperature profile of an atmosphere in radiative--convective equilibrium. This methodology has been successfully applied to modeling solar system planets and moons, brown dwarfs, and exoplanets, with both cloudy and clear atmospheres; the models are described in \ct{Mckay89, Marley96, Burrows97, MM99, Marley02, Fortney05, Saumon08, Fortney08b}. 


The atmosphere model utilizes the radiative transfer techniques described in \ct{Toon89}. Within this method, it is possible to include Mie scattering of particles as an opacity source in each layer. Our opacity database for gases, described extensively in \ct{Freedman08}, includes all the important absorbers in the atmosphere. This opacity database includes two significant updates since \ct{Freedman08}, which are described in \ct{Saumon12}: a new molecular line list for ammonia \cp{Yurchenko11} and an improved treatment of collision induced H$_2$ absorption \cp{Richard12}.

The equilibrium cloud model is coupled with the radiative transfer calculations and the pressure-temperature profile of the atmosphere; this means that a converged model will have a temperature structure that is self-consistent with the clouds. Figure \ref{pt-profile} shows an example of how clouds change the \emph{P--T} structure of an irradiated planet; the deep atmosphere of a cloudy model (dashed line) is cooler than the corresponding cloud-free model (solid line) at a given pressure in the atmosphere. This cooling is due to the opacity of the cloud, which prevents the stellar flux from warming those deep layers of the atmosphere, the so-called anti-greenhouse effect. 

The photochemical output is calculated based on a converged cloud-free model, and so does not have this same self-consistency. The opacity of the cloud is included during the \emph{P--T} structure calculation, ensuring that the atmosphere is in radiative--convective equilibrium, but a shift in the \emph{P--T} profile does not change the location of the haze layer. 

We calculate the effect of the model cloud distribution on the flux using Mie theory to describe the cloud opacity. Assuming that particles are spherical and homogeneous, we calculate the scattering and absorption coefficients of each species for each of the particle sizes within the model.

\subsection{Transmission spectrum}

The transmission spectrum model calculates the optical depths for light along the tangent path through the planet's atmosphere. The model is extensively described in \ct{Fortney03} and \ct{Shabram11}. Cloud layer cross-sections generated from the model atmosphere are treated as pure absorption, and are added to the wavelength-dependent cross-sections of the gas.

\subsection{Model grid}
\begin{figure}[t]
    \includegraphics[width=3.75in]{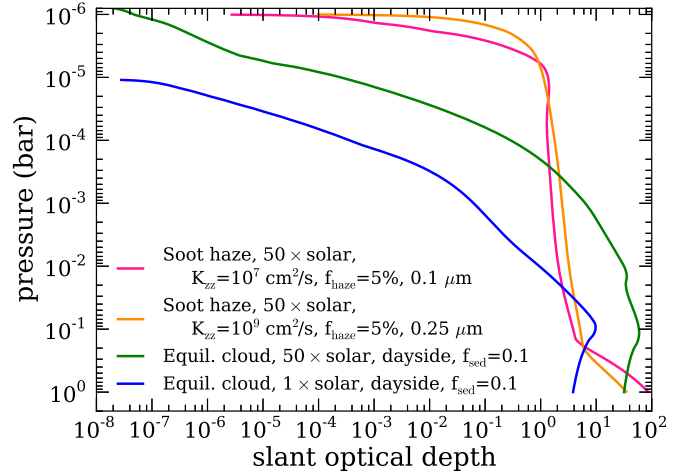}
 \caption{Slant optical depth. The slant optical depth at 1 \micron\ in four representative atmosphere models are shown. Two models include equilibrium clouds (KCl and ZnS) within the \ct{AM01} framework; the other two models include a hydrocarbon (soot) haze as described in Section \ref{hydrocarbonhaze}. The three models with enhanced (50$\times$ solar) metallicity generally match the observations (see spectra in Figures \ref{sulfspec} and \ref{sootspec4}) and have similar slant optical depths between 10$^{-3}$ and 10$^{-4}$ bar. The solar metallicity model has a lower optical depth and does not match observations.} 
\label{colopd}
 \end{figure}
 
\begin{figure*}[t]
    \includegraphics[width=7.5in]{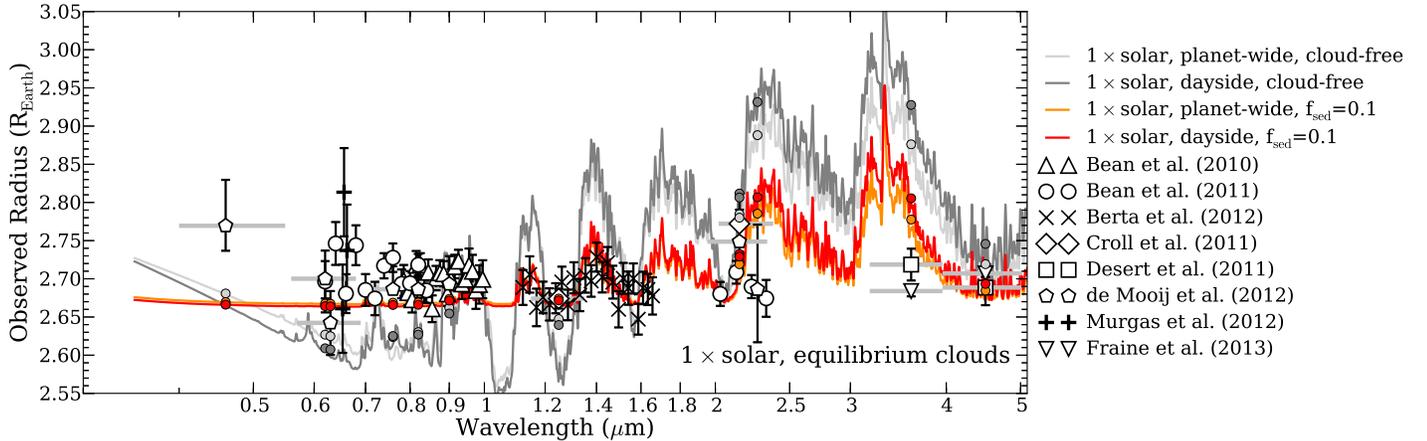}
 \caption{Reported transmission spectrum data compared to equilibrium cloud models of solar composition atmospheres. Data from a variety of sources are shown; the horizontal error bars show the width of the photometric band. Model spectra for cloud-free and cloudy solar atmospheres are plotted with corresponding model photometric points for the bands with data. We plot both `dayside' models, which assume no redistribution of heat to the nightside of the planet, and `planet-wide' models that assume that the heat is fully redistributed. Cloud-free models have features in the optical and near-IR that are inconsistent with data; cloudy models have somewhat smaller features in the near-infrared, but the features are not small enough to be consistent with the data.}
\label{sulfspecsolar}
\end{figure*}

\begin{figure*}[t]
    \includegraphics[width=7.5in]{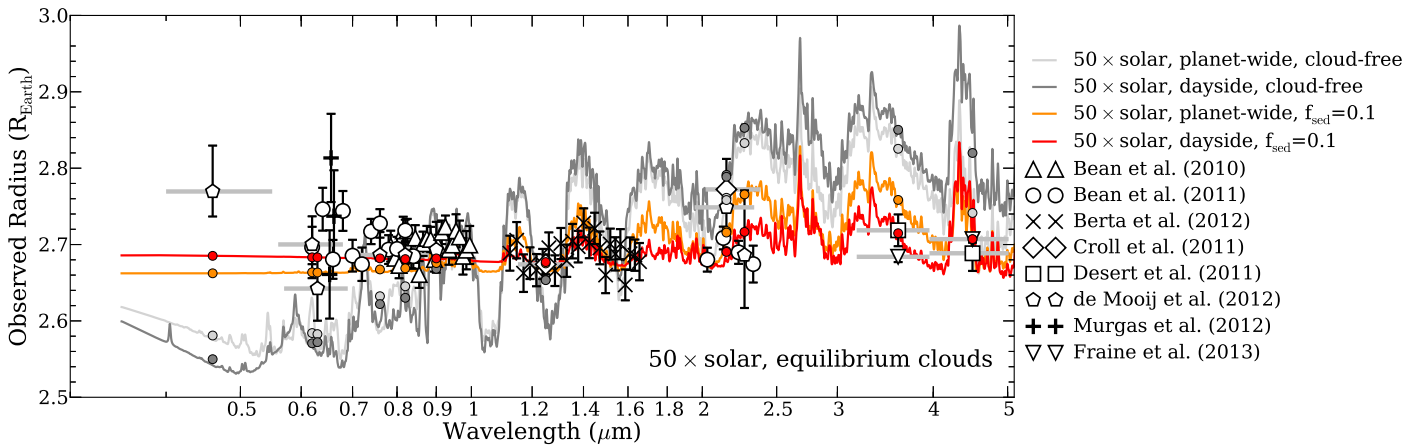}
 \caption{Reported transmission spectrum data compared to equilibrium cloud models of 50$\times$ solar composition atmospheres. Data and models are plotted as in Figure \ref{sulfspecsolar}. Cloud-free models have features in the optical and near-IR that are inconsistent with data; the cloudy 'dayside' model has a relatively flat spectrum that is generally consistent with the data.}
\label{sulfspec}
\end{figure*}

We run models with solar composition and 50$\times$ solar composition, with two different heat redistribution parameters from fully redistributed (planet-wide average) to a dayside-average. We include equilibrium clouds at a variety of different values of sedimentation efficiency \fsed: 0.1, 0.25, 0.5, and 1.0. We include a hydrocarbon haze with mode particle sizes of 0.01, 0.05, 0.1, 0.2, 0.5, 0.75, and 1.0\micron\ and soot-producing efficiencies ($f_{\rm haze}$) from 0.1 to 5\% (50$\times$ solar) and 5-25\% (solar). 

\section{Results}

We find a variety of cloudy models that are consistent with the majority of data for GJ 1214b. In general, optically thicker clouds are favored by high metallicity, efficient hydrocarbon polymerization (high $f_{haze}$), rapid vertical mixing, and more vertically extended (low \fsed) clouds with smaller particle sizes.

\subsection{Optical depths of clouds}

In order to match the observations of GJ 1214b's transmission spectrum, the cloud must be optically thick relatively high in the atmosphere (roughly 10$^{-3}$ bar), masking the strong absorption features in the infrared that would otherwise be present. Figure \ref{colopd} shows the slant optical depth at 1 \micron\ of four representative models. The slant optical depth along the terminator is a factor of $\sim$20 larger than the vertical optical depth for GJ 1214b \cp[see][]{Fortney05c}. The three enhanced-metallicity models shown become optically thick at mbar pressures; the solar composition model becomes optically thick too deep in the atmosphere to obscure the transmission spectrum.

\subsection{Equilibrium clouds}

Figures \ref{sulfspecsolar} and \ref{sulfspec} show cloud-free and cloudy model spectra that include KCl and ZnS clouds. Figure \ref{sulfspecsolar} shows solar composition models and Figure \ref{sulfspec} shows enhanced-metallicity 50$\times$ solar composition models.

Examining the solar composition model spectra (Figure \ref{sulfspecsolar}) and data by eye, the features in the infrared are larger in the models than in the data, even for \fsed$=0.1$ clouds. This suggests that if GJ 1214b does have a solar-metallicity atmosphere, these clouds alone are not likely to be fully obscuring the near-infrared spectrum. 

However, in the enhanced-metallicity models (Figure \ref{sulfspec}), the \fsed$=0.1$ models become optically thick high enough in the atmosphere to match the observations. Models with higher values of \fsed\ (i.e. thinner clouds) have features in the optical and infrared larger than the data show. Hotter models (with inefficient heat redistribution) match better than models with efficient planet-wide redistribution because the \emph{P--T} profile crosses the condensation curve at a higher altitude, forming the cloud higher in the atmosphere. For the cloudy models shown in Figure \ref{sulfspec}, the mode particle sizes calculated by the cloud model range between 0.02 \micron\ at low pressures (10$^{-6}$ to 10$^{-5}$ bar) to $\sim$10 \micron\ near the cloud base.  

\begin{figure*}[t]
  \begin{minipage}[b]{ \linewidth}
    \includegraphics[width=7.5in]{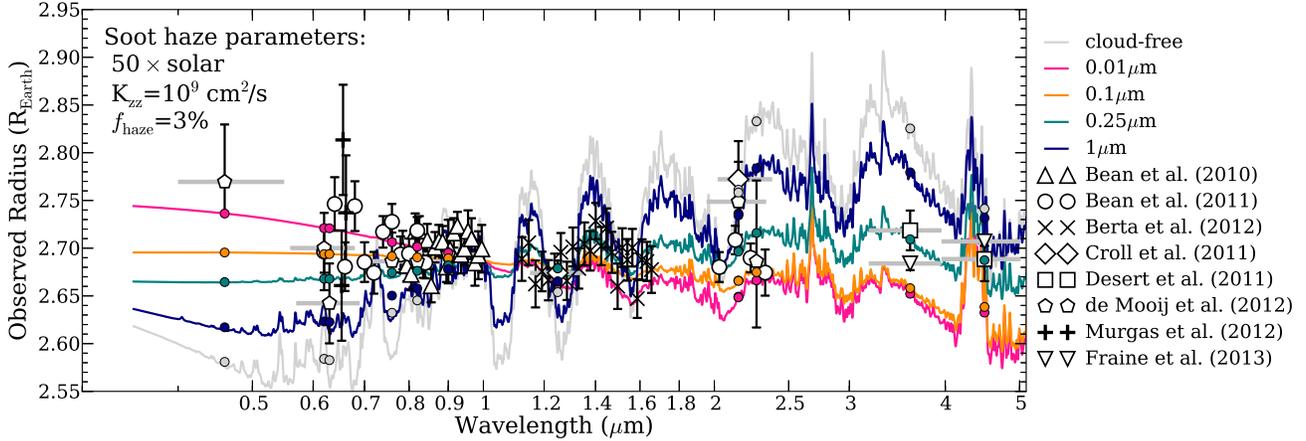}
  \end{minipage}
 \caption{The effect of particle size on the transmission spectrum is shown. Data are compared to 50$\times$ solar composition hydrocarbon haze models. Data from a variety of sources are shown; the horizontal error bars show the width of the photometric band. The model radii integrated over the photometric band are shown for each photometric data point. All models have 50$\times$ solar composition and use the photochemical results for K$_{zz}$=10$^9$ cm$^2$ s$^{-1}$ models. All models use a 3\% soot-forming efficiency ($f_{\rm haze}$) so the mass of haze particles in each layer is the same. Particle size has a strong effect on the cloud opacity. The smallest particles are the most optically thick in the optical; large particles are fairly optically thin because, given the same amount of cloud mass, their number density is significantly lower. }
\label{sootspec1}
\end{figure*}

\begin{figure*}[t]
    \includegraphics[width=7.5in]{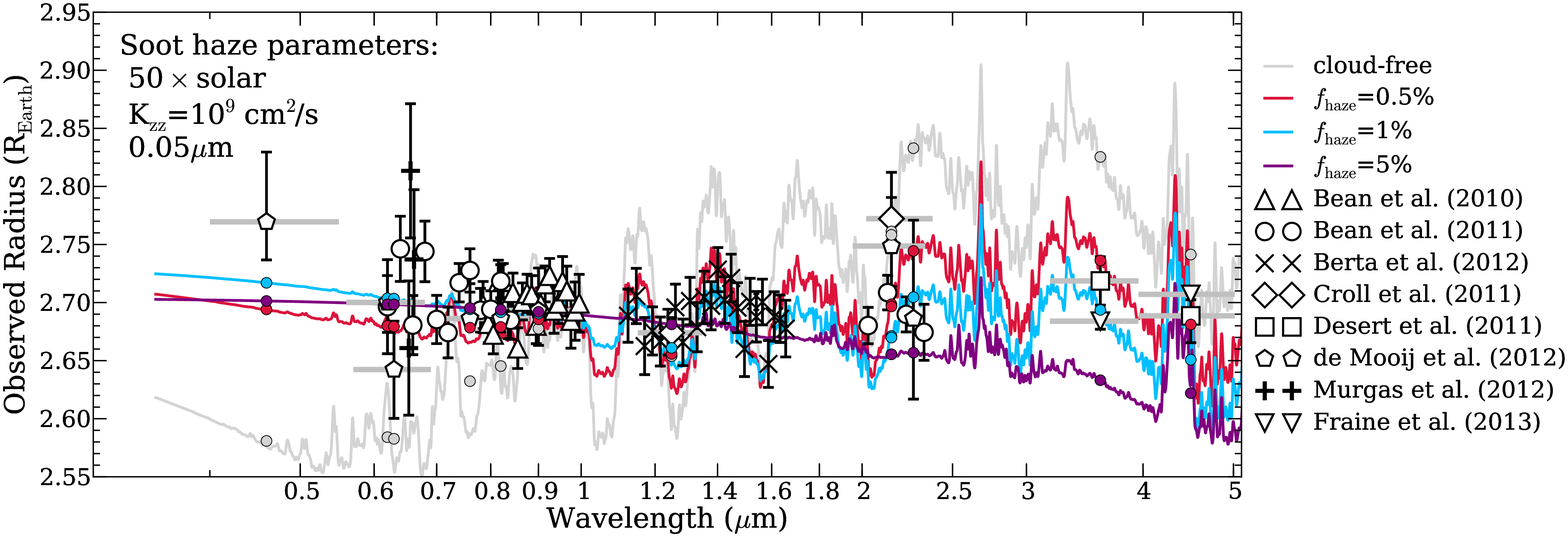}
 \caption{The effect of $f_{\rm haze}$ on the transmission spectrum is shown. Data are compared to solar composition hydrocarbon haze models. Data from a variety of sources are shown; the horizontal error bars show the width of the photometric band. The model radii integrated over the photometric band are shown for each photometric data point. All models have solar 50$\times$ solar composition, a 0.05\micron\ mode particle size, and K$_{zz}$=10$^9$ cm$^2$ s$^{-1}$.  Higher values of $f_{\rm haze}$ lead to optically thicker clouds and a more obscured transmission spectrum.}
\label{sootspec2}
\end{figure*}

\subsubsection{Chi-squared analysis}

In addition to fitting by-eye, we perform a simple chi-squared analysis to understand the validity of our fits. We tested an algorithm similar to that used in \ct{Cushing08} in which we weight by the width of the band fitted, to avoid treating spectroscopy much more heavily than photometry. We get qualitatively identical results for the best-fitting models with and without this weighting parameter, so for simplicity we present the unweighted results here. 

At solar metallicity, for cloud-free models, the reduced chi-squared (\rchi) is 37.9 and 26.2 respectively for the dayside and planet-wide models. For the cloudy \fsed$=0.1$ models, \rchi is 8.2 (dayside) and 5.8 (planet-wide). In comparison, for a 100\% water atmosphere (spectrum shown in Figure \ref{zoomspec}), \rchi\ is 1.4. In agreement with the by-eye fit, all solar composition models fit more poorly than a steam atmosphere. 

At 50$\times$ solar metallicity, \rchi\ for the cloud-free models is 17.7 (planet-wide) and 31.5 (dayside). For the cloudy \fsed$=0.1$ models, \rchi\ is 4.4 (planet-wide) and 1.9 (dayside). For all \fsed$\geq$0.2, \rchi$>5$. \fsed$=0.1$ models with partially inefficient redistribution are the only models the match the data as well as a water atmosphere. In section \ref{dis-fsed} we discuss how this sedimentation efficiency compares to brown dwarfs and whether it appears to be physically reasonable. 

\subsection{Hydrocarbon haze}

\begin{figure*}[t]
    \includegraphics[width=7.5in]{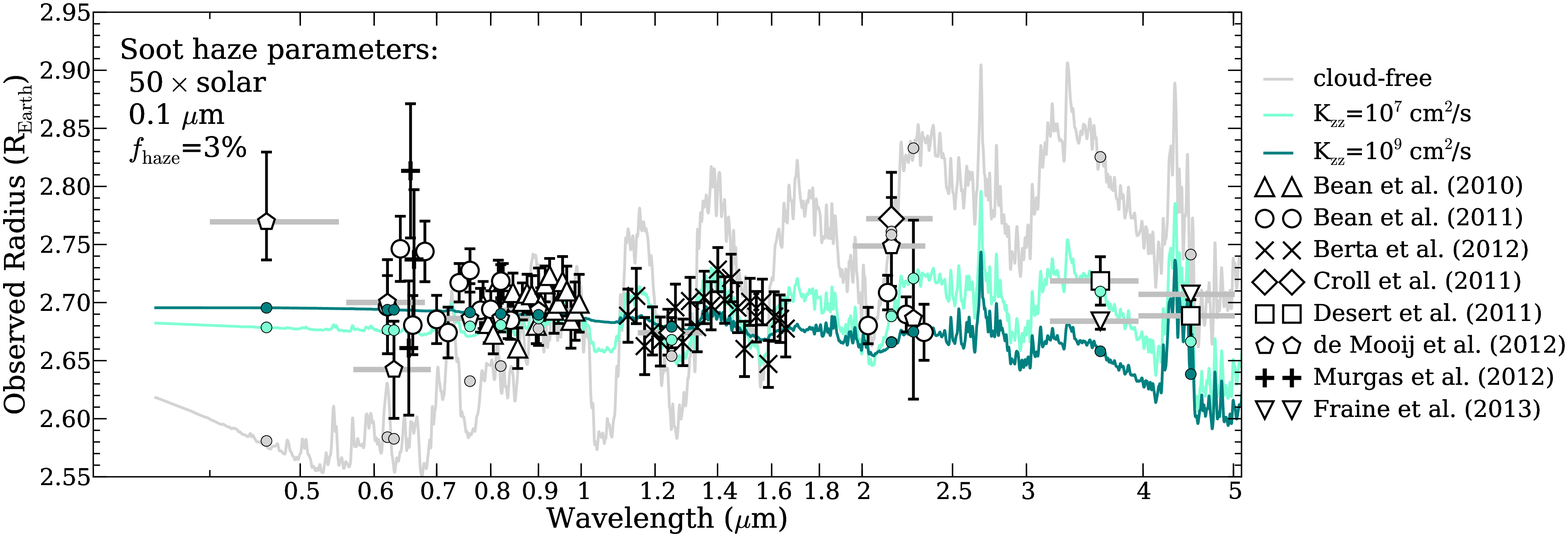}
 \caption{The effect of vertical mixing on the transmission spectrum is shown. Data are compared to solar composition hydrocarbon haze models. Data from a variety of sources are shown; the horizontal error bars show the width of the photometric band. The model radii integrated over the photometric band are shown for each photometric data point. All models have solar 50$\times$ solar composition, a 0.1$\mu$ mode particle size, and a soot-forming efficiency $f_{\rm haze}$=3\%. The eddy diffusion coefficient \kzz, which parametrizes the strength of vertical mixing, is varied between \kzz=10$^7$ to 10$^9$ cm$^2$ s$^{-1}$. \kzz has a strong effect on the cloud opacity. More vertical mixing lofts more soot-forming material high in the atmosphere; the cloud is therefore most optically thick in the near infrared for \kzz=10$^9$cm$^2$ s$^{-1}$. }
\label{sootspec3}
\end{figure*}

\begin{figure*}[t]
    \includegraphics[width=7.5in]{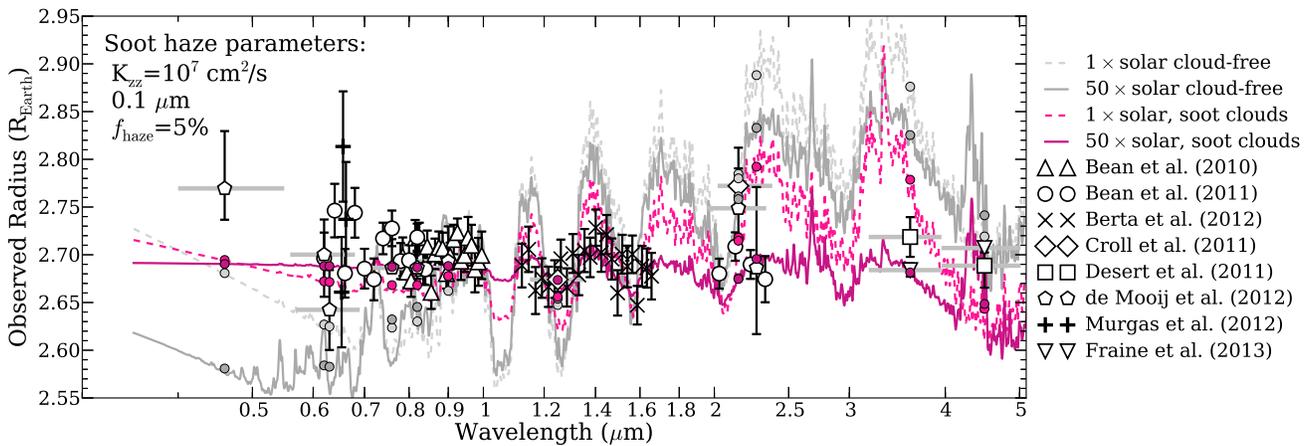}
 \caption{The effect of both metallicity and hazes on the transmission spectrum is shown. Data are compared to solar composition and 50$\times$ solar models, with and without hydrocarbon hazes. Data from a variety of sources are shown; the horizontal error bars show the width of the photometric band. The model radii integrated over the photometric band are shown for each photometric data point. All models have a 0.1$\mu$ mode particle size, and a soot-forming efficiency of 5\%. The eddy diffusion coefficient \kzz, which parametrizes the strength of vertical mixing, is \kzz=10$^7$ cm$^2$ s$^{-1}$.  Solar composition models with hazes generally are generally not flatted enough to become consistent with the data.}
\label{sootspec4}
\end{figure*}

Figures \ref{sootspec1}, \ref{sootspec2}, \ref{sootspec3}, and \ref{sootspec4} show examples of the extensive grid of models that include a hydrocarbon haze layer. In these four figures, the effects of the four parameters we vary are shown. 

Figure \ref{sootspec1} illustrates the effect of changing the mean particle size on the transmission spectrum. Each of the models shown has the same $f_{\rm haze}$ value (3\%) and uses the same photochemistry (50$\times$ solar, $K_{zz}$=10$^9$ cm$^2$ s$^{-1}$), so the mass of haze particles in each model is identical, isolating the effect of particle size. Small particles are generally more optically thick because, given the same total haze mass, smaller particles have a higher number density. For the smallest particle sizes (0.01 $\mu$m), scattering by haze particles causes the transmission spectrum to rise into the optical. For larger particles, they scatter less efficiently at optical wavelengths. For particle sizes above $\sim$0.25 $\mu$m, the opacity of the haze particles is relatively gray for optical through near-infrared wavelength and the resulting spectrum is more flat.

Figure \ref{sootspec2} illustrates the effect of changing the fraction of haze precursors that actually form into haze particles (defined here as $f_{\rm haze}$, see equation \ref{fhaze-eq}). The models shown each have the same photochemistry (50$\times$ solar, $K_{zz}$=10$^9$ cm$^2$ s$^{-1}$) and particle sizes (0.05 \micron) to isolate the effect of changing the fraction of soot precursors that become haze particles. As expected, increasing $f_{\rm haze}$ increases the optical depth of the haze, obscuring the molecular features in the spectrum.

Figure \ref{sootspec3} shows how vertical mixing, parametrized as \kzz\ in the photochemistry models, affects the transmission spectra. The same metallicity (50$\times$ solar), particle size (0.1 $\mu$m) and $f_{\rm haze}$ (3\%) are used, and \kzz\ is varied from 10$^7$--10$^9$ cm$^2$ s$^{-1}$. Generally, we find that the eddy diffusion coefficient \kzz\ affects the haze-forming efficiency needed to reproduce the observations. The stronger the vertical mixing, the larger the quantity of soot precursors (see also Figure \ref{photochem}). This means that models with $K_{zz}$=10$^9$ cm$^2$ s$^{-1}$ have more optically thick haze than $K_{zz}$=10$^7$ cm$^2$ s$^{-1}$; if vertical mixing is more efficient, a lower fraction of soot precursors need to form into haze material to match the observations. 

Figure \ref{sootspec4} shows the effect of metallicity on the model transmission spectra. Unsurprisingly, we find that the 50$\times$ solar metallicity models have significantly more soot precursors (see Figure \ref{photochem}), as there are a factor of 50 more heavy elements in the atmosphere. This means that if the same fraction of soot precursors become haze particles, the high metallicity model will have more mass in haze and therefore a more optically thick atmosphere. Indeed, we find that very few of the models at solar metallicity have an optically thick haze layer. 

\subsubsection{Best-fitting hydrocarbon haze models}
 
\begin{figure*}[t]
  \begin{minipage}[b]{ 0.5 \linewidth}
     \hspace{-0.65in} \includegraphics[width=4.5in]{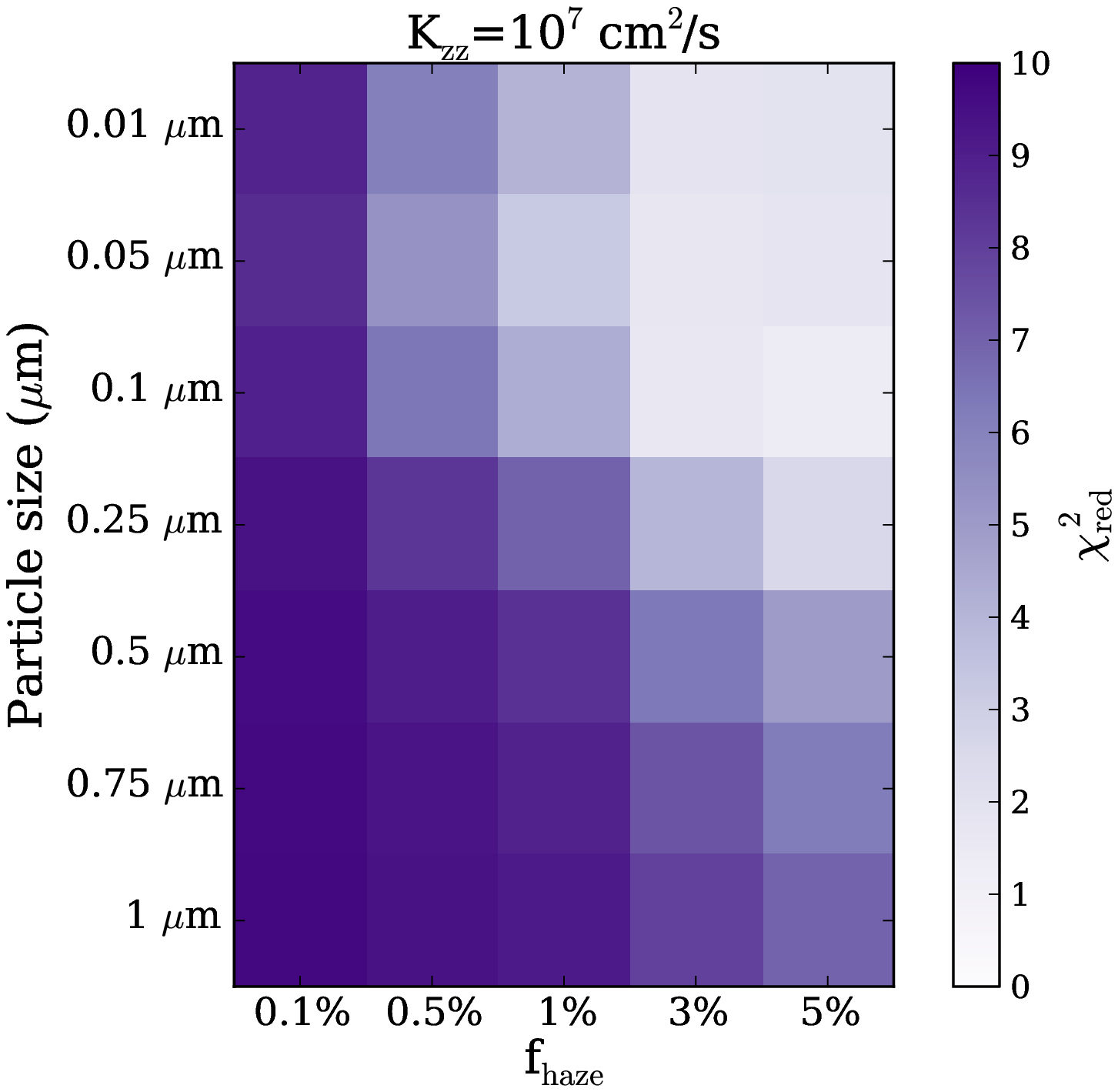}
  \end{minipage}
  \begin{minipage}[b]{ 0.5 \linewidth}
     \hspace{-0.65in} \includegraphics[width=4.5in]{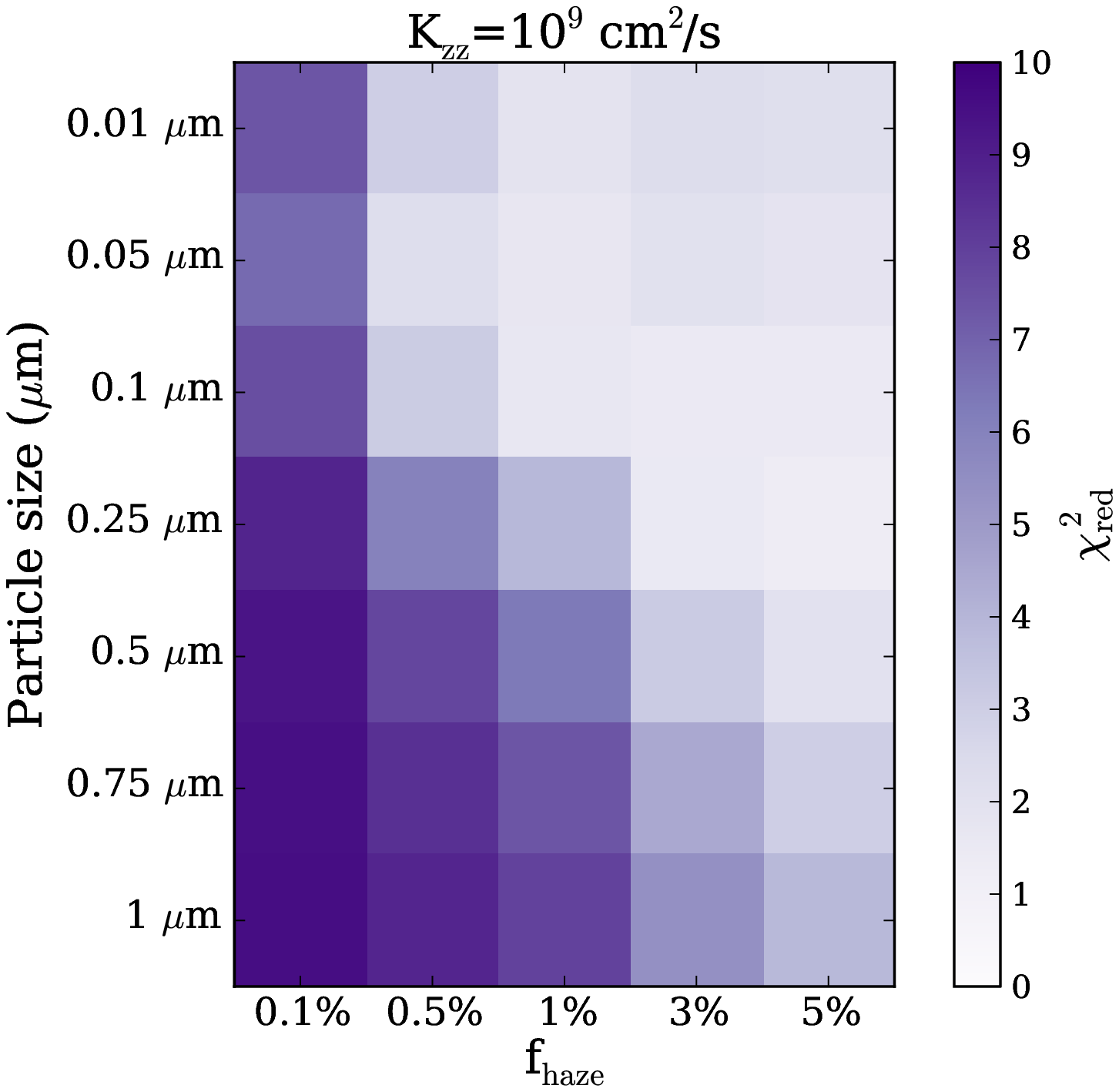}
  \end{minipage}  
 \caption{$\chi_{\rm red}^2$ for 50$\times$ solar models with hazes. The goodness-of-fit parameter $\chi_{\rm red}^2$ for each of the 50$\times$ solar hydrocarbon haze models is plotted. \kzz=10$^7$ cm$^2$ s$^{-1}$ is on the left and \kzz=10$^9$ cm$^2$ s$^{-1}$ is on the right. At each particle size and $f_{\rm haze}$ value, the shading indicates the goodness of the fit with lighter shades indicating a better fit. It is clear that small particles and moderate to high $f_{\rm haze}$ is necessary to reproduce the majority of the observed transmission spectrum. The range of well-fitting models is larger for the more vigorous (\kzz=10$^9$ cm$^2$ s$^{-1}$) vertical mixing. }
\label{chisquaredplot}
\end{figure*}

In general, we find a range of models with a hydrocarbon haze layer that can match most of the observations. The best-fitting models all have 50$\times$ solar metallicity. For the less vigorous mixing (\kzz=10$^7$ cm$^2$ s$^{-1}$), models with small particle sizes (0.01 to 0.1 \micron) and $f_{\rm haze}$ of 3-5\% match the data; for more vigorous mixing (\kzz=10$^9$ cm$^2$ s$^{-1}$), models with small particles (0.01 to 0.1 \micron) and $f_{\rm haze}$ of 1-5\% match the data, as do medium-sized particles (0.25 \micron) with $f_{\rm haze}$ from 3-5\%. This parameter space of models is summarized in Figure \ref{chisquaredplot}, which shows the well-fitting parameter space as light shaded regions and the poor-fitting parameter space as darker shaded regions. 

At solar metallicity, a very small subset of the parameter space resulted moderately well-fitting models. No models with solar metallicity and \kzz=10$^7$ cm$^2$ s$^{-1}$ had a $\chi_{\rm red}^2$ less than 4. For the more vigorous  \kzz=10$^9$ cm$^2$ s$^{-1}$, only a single model had a reasonably good fit ($\chi_{\rm red}^2$=3), which had particle sizes of 0.25 \micron\ and $f_{\rm haze}$=25\%. This $f_{\rm haze}$ value represents a quarter of soot precursors forming into condensed haze solids, which seems quite high. 

These results generally suggest that if GJ 1214b has an enhanced metallicity atmosphere like Neptune, there is a large range of particle size distributions and photochemical efficiencies that can result in an obscuring haze in the atmosphere.

\subsection{Combinations of cloud layers}

In a planetary atmosphere, a number of different cloud and haze layers can form. For example, in Titan's atmosphere, there is both a high photochemical hydrocarbon haze and a deeper methane cloud. To examine this for GJ 1214b, we include both the equilibrium KCl and ZnS clouds and the hydrocarbon soot layer in a set of solar composition models, to see if by including both clouds we could match the spectrum without enhancing the metallicity of the atmosphere.

We ran a small set of models with favorable equilibrium cloud parameters (no heat redistribution to the night side, \fsed=0.1) and hydrocarbon haze parameters ($f_{\rm haze}$=5-10\%, \kzz=10$^9$ cm$^2$ s$^{-1}$). However, none of these models fit the data as well as the enhanced-metallicity equilibrium cloud models, enhanced-metallicity hydrocarbon haze models, or a high mean molecular weight water-rich model.

\begin{figure*}[t]
  \begin{minipage}[b]{ \linewidth}
  \center    \includegraphics[width=7in]{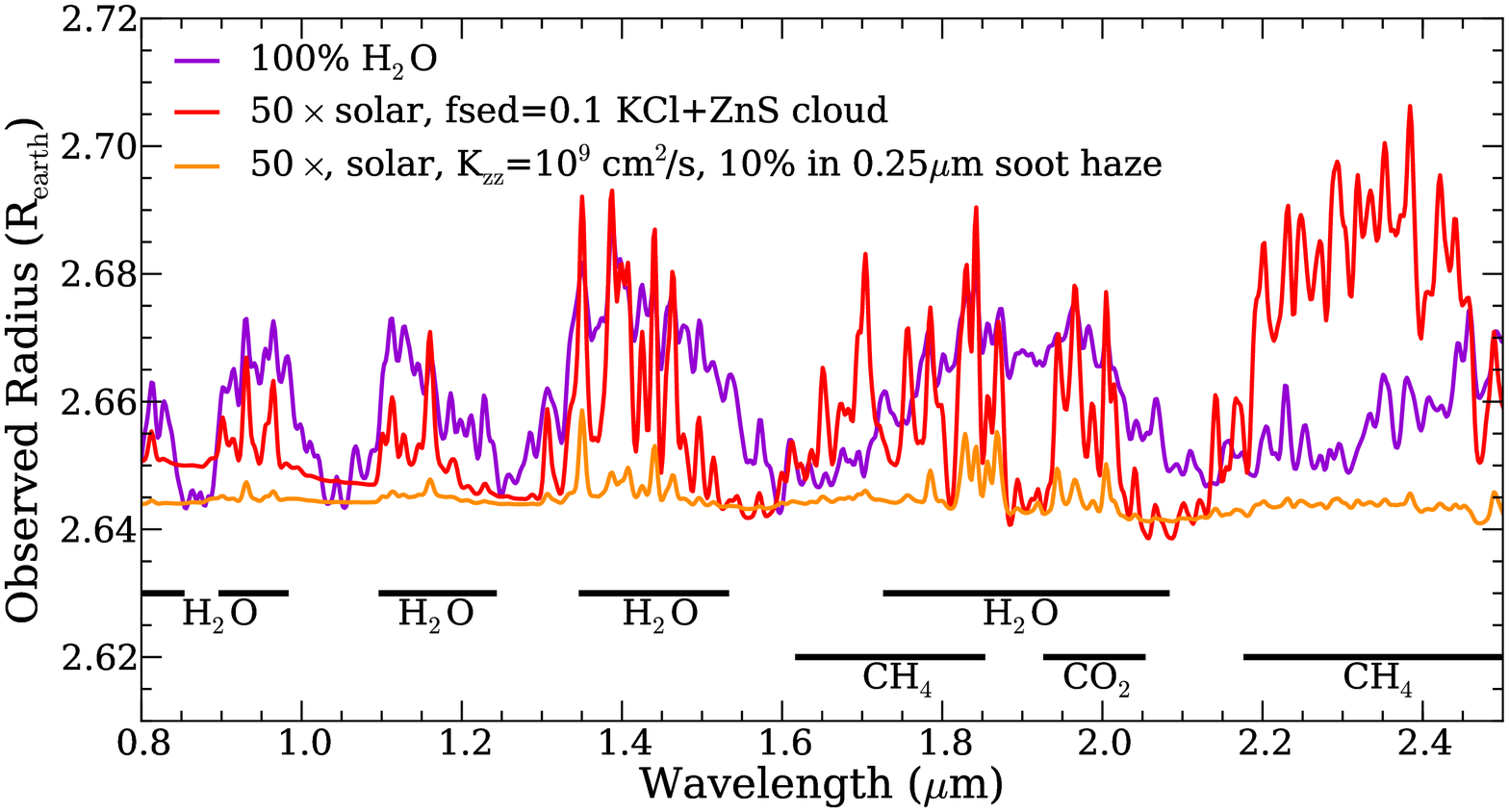}
  \end{minipage}
 \caption{Comparison of steam and cloudy H-rich atmosphere models. A 100\% water atmosphere is compared to two cloudy H-rich models in the near-infrared. With a higher-fidelity near-infrared spectrum, these models could be easily distinguished. Locations of strong absorption features from H$_2$O, CH$_4$, and CO$_2$ are noted. The Hubble Space Telescope G141 grism has a maximum resolving power of 130 in the range 1.1--1.7 \micron.}
\label{zoomspec}
\end{figure*}

\begin{figure*}[t]
  \begin{minipage}[b]{ \linewidth}
  \center    \includegraphics[width=7in]{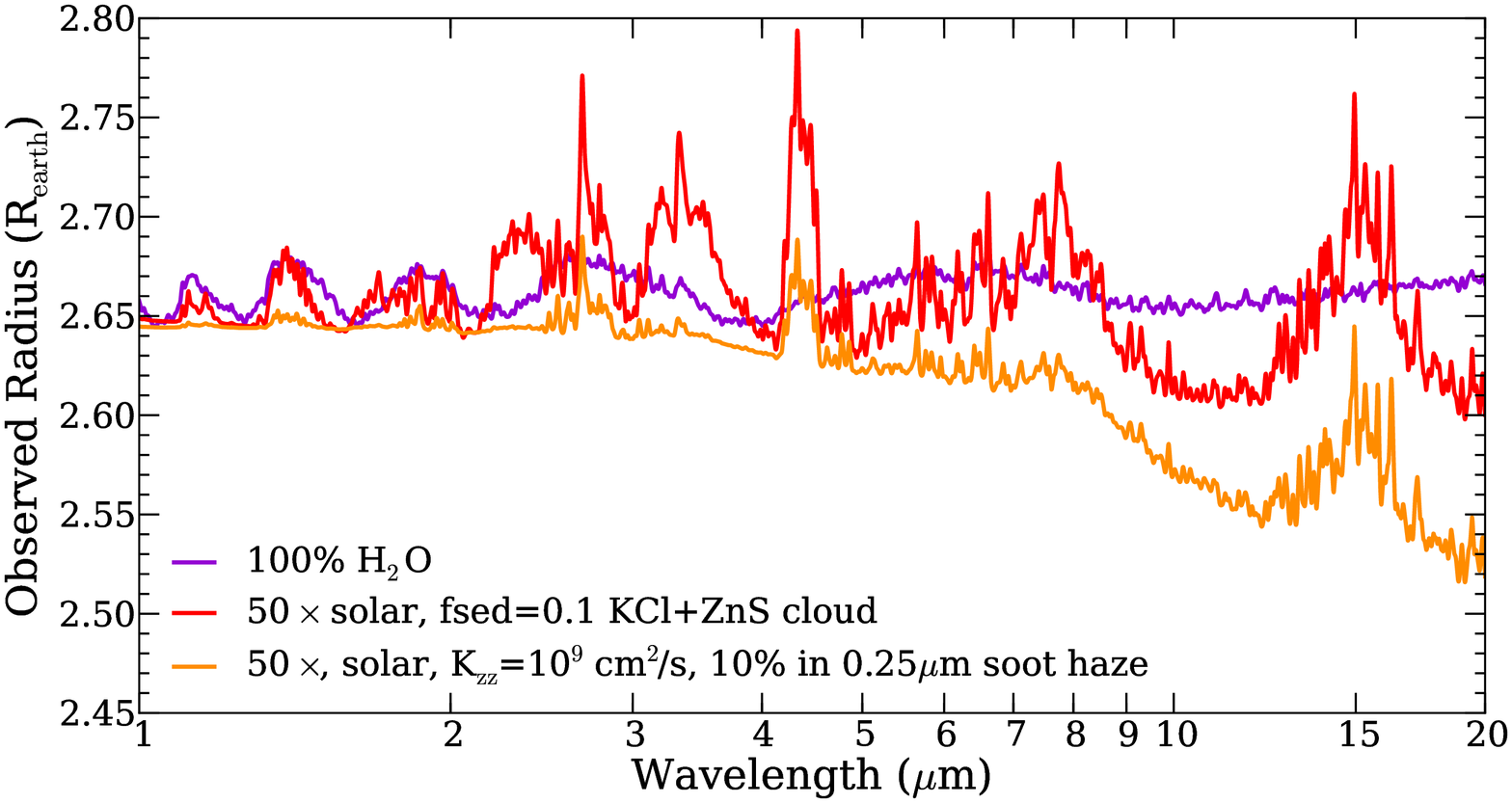}
  \end{minipage}
 \caption{Comparison of steam and cloudy H-rich atmosphere models in the mid-infrared. The models from Figure \ref{zoomspec} are shown for a wider wavelength range. The 100\% water atmosphere model shows water vapor features of a similar amplitude from 1--20\micron. However, for both of the cloudy models, the clouds become significantly less optically thick at longer wavelengths than they are in the near-infrared where current data exists. This means that in the mid-infrared, the features are much larger. }
\label{zoomspec20micron}
\end{figure*}

\section{Discussion}

\subsection{Physical nature of low \fsed\ values} \label{dis-fsed}

As discussed in Section \ref{methods-eqclouds}, the particle sizes and vertical thickness of the equilibrium KCl and ZnS clouds are calculated using the parametrized value \fsed, which is equal to the ratio of the sedimentation velocity to the updraft velocity. A high sedimentation efficiency \fsed\ forms a cloud with large particles that settles into a thin layer; a low \fsed\ forms a more extended cloud with small particles. 

This model has been used most frequently for studies of brown dwarfs. Studies of L dwarfs find that \fsed$\sim$1--3 for the majority of field L dwarfs \cp{Stephens09, Saumon08}. Similarly, \ct{Morley12} found that for sulfide clouds in T dwarfs, \fsed$\sim$4--5. 

In this study, we find that the value needed to fit the observations is \fsed$=0.1$, a sedimentation efficiency more than ten times lower than those of brown dwarfs. However, this low value may not be unreasonable for an irradiated planetary atmosphere. In \ct{AM01}, values of \fsed\ for Earth clouds are calculated. They find that for clouds that form high in Earth's atmosphere---stratocumulus clouds---\fsed\ is less than 1, with values for specific case studies ranging from 0.2 (North Sea) to 0.3--0.5 (California). The clouds we model in GJ 1214b form within a nearly-isothermal radiative region of the atmosphere, so we expect them to behave more like Earth stratocumulus clouds than the deeper tropospheric cumulus clouds, which have high \fsed\ (2--6), more similar to brown dwarfs. 

\fsed\ is the ratio of the sedimentation velocity to the updraft velocity which is equal to \kzz$/L$, where $L$ is the mixing length. A low \fsed\ could be caused by many different things. (1) If the cloud particles are fluffy aggregates, they would have a slow sedimentation velocity. (2) If \kzz\ is large, like those of hot Jupiters (\ct{Showman09} finds that hot Jupiters have \kzz$\sim10^{11}$cm$^2$ s$^{-1}$), then the updraft velocity will be large. (3) If the mixing length $L$ is small (due to a mean molecular weight gradient or wave breaking effects), then the atmosphere will be stably stratified, and \fsed\ will be small.

There is already clear observational evidence of a cloud layer very high in the atmosphere of hot Jupiter HD 189733 \cp{Pont08,Sing11,Pont12}. The spectral slope of the observations suggests opacity due to Rayleigh scattering, which would be due to quite small (sub-micron) sized particles. The cloud layer obscures gaseous absorption features from the blue to the near infrared. Based on its optical properties, the obscuring cloud layer has been suggested to be composed on small enstatite particles \cp{Lecavelier08b}, which would have to be kept aloft high in the planet's atmosphere due to inefficient sedimentation.

\subsection{Distinguishing between a steam and cloudy atmosphere}

GJ 1214b's transmission spectrum is often described as `flat' or `featureless,' but in reality, its features are just too small to detect with current signal-to-noise observations. While we find that with current data, a hydrogen-helium rich model with clouds can fit just as well as a 100\% water model, if we can improve the precision in the near-infrared, there are features that allow us to distinguish between these possibilities \cp[see also][] {Benneke12}. 

Figure \ref{zoomspec} shows how two sample cloudy models compare to a 100\% water model. The hydrocarbon haze model is deliberately chosen to show the large extent to which the haze layer can obscure the near-infrared features and flatten the spectrum. One feature of cloudy spectra, absent in the 100\% water spectrum, is that there are flat regions between features, especially at 0.9, 1.1, and 1.3 \micron. This is the pressure level where the clouds become optically thick; above this level, one can see gas opacity features, but all features below are obscured. A higher signal-to-noise spectrum in the near-infrared from 1 to 1.8 \micron\ should be able to distinguish between these possibilities. 

The spectra also look different in regions where additional species absorb more strongly than water vapor. For example, between 2.2 and 2.4 \micron\ there is a strong methane band. (Note that this part of the spectrum has particularly conflicting results to date). Similarly, the feature at 1.7 \micron\ is due to methane and the feature at 2.0 \micron\ is due to CO$_2$. These bands would be completely lacking in a pure water atmosphere. By resolving regions of the spectrum where additional absorbers, if they exist, dominate, we could differentiate between a pure water atmosphere and a hydrogen/helium-rich atmosphere with many absorbers including clouds. 

Figure \ref{zoomspec20micron} shows the same models as Figure \ref{zoomspec}, but for longer wavelengths (1--20\micron). For both cloudy models shown, because the cloud particles are relatively small they do not absorb as efficiently in the mid-infrared as they do in the near-infrared. The cloud opacity decreases significantly, and even in models with a thick obscuring haze layer in the near-infrared, the atmosphere becomes clear of haze at mid-infrared wavelengths. Gaseous water and methane features dominate the transmission spectrum beyond 3--4 \micron.  Promisingly, this suggests that even if many exoplanet atmospheres' features are obscured in the wavelengths accessible from the ground or from \emph{HST}, the wavelengths probed by the \emph{James Webb Space Telescope} will be less sensitive to haze obscuration. 

However, atmospheres with mixed H$_2$O/H$_2$-rich compositions can also fit the data; \ct{Berta12} finds that any atmosphere with a mass fraction of water higher than 70\% can fit the observations. Distinguishing a mixed atmosphere from a cloudy H$_2$-rich atmosphere would be more challenging as features such as methane may also appear. The chemistry of such extremely high metallicity atmospheres is not currently well-understood and is a subject of ongoing study \cp{Moses12b}. 

\subsection{Photochemical processes}

This work, and that of \ct{Zahnle09} and \ct{Kempton12}, suggests that photochemistry could be extremely important for interpreting the spectra of cool exoplanets. We find that there is a large range of parameter space for a photochemical haze that can obscure the transmission spectrum of a hydrogen and helium dominated atmosphere. In this work, we parametrized the chemical processes expected to polymerize 2nd-order hydrocarbons. As part of our ongoing and future work we seek to identify and characterize the key chemical pathways expected to produce higher-order hydrocarbons in the upper atmospheres of irradiated exoplanets. 

The spectral signatures of photochemically-produced gases were not included in these spectra, but it has been suggested that these would have relatively small signatures \cp{Kempton12}. Detecting additional soot precursors like benzene rings, polycyclic aromatic hydrocarbons, or other polymers with high-resolution spectroscopy would further constrain photochemical haze creation. 

\subsection{C/O ratio}
\label{coratio}
Recent work has shown that planets may have some range in carbon and oxygen abundances \cp{Madhu11b, Madhu11, Madhu12b, Moses13}. In particular, \ct{Moses13} studied the effect of C/O ratio on disequilibrium processes such as photochemistry and vertical mixing. They find that in atmospheres with a high C/O ratio, the abundances of soot precursors such as HCN and C$_2$H$_2$ are significantly enhanced. If GJ 1214b did have a high C/O ratio, it may be even easier to form a layer of optically thick soot. 



\section{Conclusions}

Previous work by \ct{Howe12} has shown that by adding an ad-hoc haze layer, the observations of GJ 1214b can be reproduced. Here, we showed that two types of clouds that may naturally emerge from equilibrium or non-equilibrium chemistry considerations, in an enhanced-metallicity atmosphere, can reproduce the observations of GJ 1214b. We presented results that show that clouds that form as a result of equilibrium chemistry, as they perhaps do on brown dwarfs, can reproduce the observations of GJ 1214b if they are lofted high in the atmosphere and the sedimentation efficiency parameter \fsed\ is low (0.1). This value is significantly different than the values of \fsed$\sim$1--3 for L dwarfs or $\sim$4--5 for T dwarfs, but is potentially quite reasonable for high altitude clouds in an irradiated planet.

We showed that models including hydrocarbon haze that forms as a result of photochemistry can also flatten GJ 1214b's spectrum. We used a 1D photochemical kinetics model to calculate the vertical distribution and available mass of molecules that are produced on the pathway to haze formation.  With haze-forming efficiencies between 1\% and 5\%, we found equally well-fitting models with modal particle sizes from 0.01 to 0.25\micron. We conclude that, while more work on understanding the chemical processes of forming hydrocarbons is necessary, it is very plausible that GJ 1214b's spectrum is obscured by a layer of soot. 

Although there are of course uncertainties in the detailed implementation of the cloud models, we stress that both kinds of clouds emerge naturally from either equilibrium chemistry or photochemical arguments.  In particular, haze formation has the possibility to lead to the obscuration of gaseous absorption features over a wide range of planetary parameter space, from super-Earths to giant planets, over a wide range in planetary temperature.

\acknowledgements
This work benefited from helpful conversations with Jacob Bean and support from NASA grants HST-GO-12251.05-A and NNX12AI43A. MSM acknowledges support of NASA PATM program. CV acknowledges support from NASA PATM grant NNX11AD64G. 

\vspace{0mm}

\bibliographystyle{apj}

\end{document}